\UseRawInputEncoding
\documentclass[a4paper,aps,prd,10pt,preprintnumbers,showpacs,
twocolumn,superscriptaddress,nofootinbib,amsmath,amssymb]{revtex4-1}
\usepackage{graphicx}
\usepackage{cmap}
\usepackage[utf8]{inputenc}
\usepackage[T1]{fontenc}

\def\imo{i}

\def\K{{\cal K}}
\def\Order#1{{\cal O}\left(#1\right)}

\begin{document}
\title{Massive scalar quasinormal modes of an asymptotically flat regular black hole supported by a phantom Dirac--Born--Infeld field}
\author{Milena Skvortsova}\email{milenas577@mail.ru}
\affiliation{Academic Research Institute of Gravitation and Cosmology, Peoples' Friendship University of Russia (RUDN University), 6 Miklukho-Maklaya Street, Moscow 117198, Russian Federation}
\begin{abstract}
We study the quasinormal spectrum of a massive test scalar field in the exact asymptotically flat regular black-hole geometry supported by a phantom Dirac--Born--Infeld scalar. Using high-order WKB approximation improved by Pad\'e resummation, together with characteristic time-domain integration and Prony extraction, we compute the fundamental mode and the first two overtones for representative values of the regularity parameter and the field mass. We show that increasing the field mass raises the oscillation frequency and reduces the damping rate, while increasing the regularity scale generally makes the ringing softer and longer lived. The time-domain profiles are in very good agreement with the WKB--Pad\'e results and confirm the robustness of the spectrum. For sufficiently large field mass, the damping tends to zero, indicating the onset of quasiresonant behavior, although in the time domain these modes are eventually masked by oscillatory late-time tails. Our results show that massive scalar ringing provides a sensitive probe of this DBI-supported regular geometry.
\end{abstract}
\maketitle
\section{Introduction}

The ringdown stage of a perturbed compact object is governed by quasinormal modes, whose complex frequencies encode the response of the underlying geometry and matter content. For black holes, these damped oscillations provide a natural bridge between exact solutions, stability analysis, and observational black-hole spectroscopy, making them one of the most informative probes of departures from the standard Schwarzschild paradigm \cite{Kokkotas:1999bd,Berti:2009kk,Bolokhov:2025uxz}.

Among the various perturbative sectors, massive fields are especially informative because the mass term qualitatively modifies both the quasinormal spectrum and the late-time signal. Effective masses arise in several physically motivated settings: they can be induced on the brane by higher-dimensional bulk effects \cite{Seahra:2004fg}, appear in massive-gravity inspired contexts relevant for very long gravitational waves \cite{Konoplya:2023fmh,NANOGrav:2023hvm}, or emerge for otherwise massless fields in magnetized black-hole backgrounds \cite{Konoplya:2008hj,Wu:2015fwa}. A particularly striking consequence is the possibility of quasiresonances, that is, arbitrarily long-lived modes appearing for specific values of the field mass \cite{Ohashi:2004wr,Konoplya:2004wg}. This behavior has been observed for different spins and in a broad variety of black-hole and compact-object spacetimes \cite{Konoplya:2005hr,Konoplya:2017tvu,Konoplya:2018qov,Zhidenko:2006rs,Aragon:2020teq,Gonzalez:2022upu,Burikham:2017gdm,Fernandes:2021qvr,Percival:2020skc,Zinhailo:2018ska,Churilova:2020bql,Bolokhov:2023bwm,Bolokhov:2023ruj,Bolokhov:2024bke,Bolokhov:2026dzn,Lutfuoglu:2025kqp,Lutfuoglu:2025eik,Lutfuoglu:2025qkt,Lutfuoglu:2025bsf,Konoplya:2007zx,Lutfuoglu:2025hwh,Lutfuoglu:2025hjy,Lutfuoglu:2026uzy,Lutfuoglu:2026gis,Lutfuoglu:2026xlo,Lutfuoglu:2026fpx,Skvortsova:2024eqi,Skvortsova:2025cah,Skvortsova:2026unq,Dubinsky:2025wns,Dubinsky:2025bvf,Malik:2025czt}. The mass term also changes the asymptotic relaxation pattern, replacing the familiar power-law behavior by oscillatory late-time tails \cite{Jing:2004zb,Koyama:2001qw,Moderski:2001tk,Rogatko:2007zz,Koyama:2001ee,Koyama:2000hj,Gibbons:2008gg,Gibbons:2008rs,Dubinsky:2024jqi}. At the same time, long-lived massive modes are not universal, and there are examples in which they do not develop at all \cite{Zinhailo:2024jzt,Konoplya:2005hr}. These features make massive-field quasinormal modes a sensitive diagnostic of the geometry.

Regular black holes supply an equally important arena for such tests. The singularity of the Schwarzschild solution is widely regarded as a signal that classical general relativity requires an ultraviolet completion, and many singularity-resolving scenarios have therefore been explored, including non-linear electrodynamics, asymptotically safe gravity, higher-curvature corrections, and phenomenological effective descriptions. Their perturbations and quasinormal spectra have been investigated in a large number of papers \cite{Konoplya:2022hll,Fernando:2012yw,Guo:2024jhg,Huang:2023aet,Jawad:2020hju,Gingrich:2024tuf,Li:2014fka,Al-Badawi:2023lke,Cai:2021ele,MahdavianYekta:2019pol,Yang:2021cvh,Zhang:2024nny,Jusufi:2020odz,Flachi:2012nv,Lopez:2022uie}. Within this class, the Hayward geometry \cite{Hayward:2005gi} has become a representative benchmark, and its quasinormal spectrum has been studied in a number of works \cite{DuttaRoy:2022ytr,Al-Badawi:2023lke,Pedraza:2021hzw,Lin:2013ofa}. The same metric also appears, up to a redefinition of parameters, in effective-field-theory and asymptotically safe frameworks \cite{Mukohyama:2023xyf,Konoplya:2023ppx,Held:2019xde}, which illustrates how quasinormal modes of regular geometries may help distinguish between different microscopic interpretations of one and the same spacetime.

Recently, Parvez and Shankaranarayanan found an exact asymptotically flat, nonsingular black-hole solution of Einstein gravity sourced by a phantom Dirac--Born--Infeld scalar field \cite{Parvez:2025dbi}. This model is especially appealing because the geometry is not introduced phenomenologically: it follows from a definite matter action, carries scalar hair, replaces the central singularity by a regular two-sphere of finite area, and allows black-hole, extremal-remnant, and horizonless regular configurations depending on the value of the parameters \cite{Parvez:2025dbi}. The same work also emphasized possible connections with dark matter, evaporation end states, and the stability of the spacetime against massless scalar perturbations \cite{Parvez:2025dbi}. Altogether, this makes the solution a natural laboratory in which to investigate how asymptotic flatness, regularity, and a non-linear scalar source reshape black-hole ringing.

Motivated by these developments, in this paper we study the quasinormal spectrum of a massive scalar field in the above regular black-hole background. Our goal is to clarify how the field mass interacts with the regularity scale and with the specific structure of the DBI-supported geometry, and thereby extend the analytic study of quasinormal modes to a new asymptotically flat regular spacetime.

The remainder of this paper is organized as follows. In Sec.~II we review the asymptotically flat regular black-hole solution supported by the phantom DBI scalar and derive the effective wave equation for massive scalar perturbations. In Sec.~III we summarize the WKB--Pad\'e approach, the characteristic time-domain integration, and the Prony extraction procedure used in our analysis. In Sec.~IV we present and interpret the quasinormal spectrum, including its dependence on the parameters and the approach to quasiresonant behavior. Finally, in Sec.~V we summarize the main results and outline possible directions for future work.

\section{Background geometry and scalar perturbations}\label{sec:wavelike}

Following Ref.~\cite{Parvez:2025dbi}, we consider Einstein gravity minimally coupled to a non-linear Dirac--Born--Infeld scalar field. A convenient form of the action is
\begin{equation}\label{action}
S=\int d^4x\sqrt{-g}\left[\frac{R}{16\pi G}-\epsilon\Lambda^4\left(\sqrt{1+\frac{\nabla_\mu\phi\nabla^\mu\phi}{\Lambda^4}}-1\right)\right],
\end{equation}
where $\Lambda$ sets the scale of the DBI non-linearity and $\epsilon=+1$ and $-1$ correspond to the canonical and phantom branches, respectively. In the small-gradient limit the scalar sector reduces to
\begin{equation}
\mathcal{L}_{\rm DBI}=-\frac{\epsilon}{2}\nabla_\mu\phi\nabla^\mu\phi+\Order{\frac{(\nabla\phi)^4}{\Lambda^4}},
\end{equation}
so the sign of $\epsilon$ determines the effective sign of the kinetic term.

Varying the action yields Einstein's equations and the scalar equation of motion,
\begin{eqnarray}
G_{\mu\nu}&=&8\pi G\,T_{\mu\nu},\label{EinsteinEq}\\
\nabla_\mu\left(\frac{\nabla^\mu\phi}{\sqrt{1+\nabla_\alpha\phi\nabla^\alpha\phi/\Lambda^4}}\right)&=&0,\label{DBIscalar}
\end{eqnarray}
with the stress-energy tensor
\begin{equation}
\begin{split}
T_{\mu\nu}&=\epsilon\frac{\nabla_\mu\phi\nabla_\nu\phi}
{\sqrt{1+\nabla_\alpha\phi\nabla^\alpha\phi/\Lambda^4}}\\
&\quad-g_{\mu\nu}\epsilon\Lambda^4\left(\sqrt{1+\frac{\nabla_\alpha\phi\nabla^\alpha\phi}{\Lambda^4}}-1\right).
\end{split}
\end{equation}
The construction of Ref.~\cite{Parvez:2025dbi} starts from a general static, spherically symmetric ansatz and imposes the regular areal radius
\begin{equation}
R^2(r)=r^2+a^2,
\end{equation}
so that the center is replaced by a minimal two-sphere of radius $a$. A key point of that approach is that one of the gravitational field equations becomes independent of the scalar profile, allowing the metric to be determined first from regularity and asymptotic flatness. Matching the large-distance behavior to the ADM mass $M$ leads to the exact asymptotically flat geometry
\begin{equation}\label{metric}
  ds^2=-f(r)dt^2+\frac{dr^2}{f(r)}+R^2(r)(d\theta^2+\sin^2\theta d\phi^2),
\end{equation}
where
\begin{equation}\label{metric-functions}
\begin{aligned}
f(r)&=1+\frac{3M}{a}\left(\frac{r}{a}-\frac{a^2+r^2}{a^2}\arctan\frac{a}{r}\right),\\
R^2(r)&=a^2+r^2.
\end{aligned}
\end{equation}
Here $a$ is the regularity scale. At large distances one has $R(r)\sim r$ and $f(r)=1-2M/r+\Order{r^{-3}}$, while near $r=0$ the areal radius remains finite, reflecting the regular core. The remaining field equations determine the DBI scalar configuration, and the solution relevant for the present study is supported by the phantom branch discussed in Ref.~\cite{Parvez:2025dbi}. Throughout the paper we measure all dimensional quantities in units of the mass and set $M=1$.

\begin{figure}
\centering
\includegraphics[width=\linewidth]{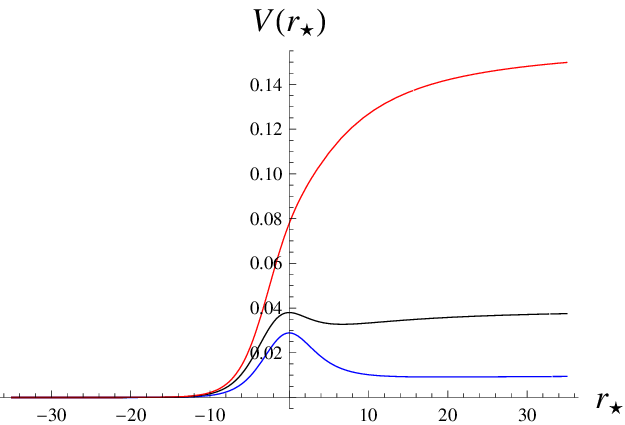}
\caption{Effective potential $V(r_*)$ for massive scalar perturbations of the asymptotically flat regular black hole with $\ell=0$, $a=0.2$, and $M=1$. The curves correspond to $\mu=0.1$ (blue), $\mu=0.2$ (black), and $\mu=0.4$ (red). As the field mass increases, the asymptotic plateau $V(\infty)=\mu^2$ rises and the barrier becomes less pronounced.}\label{fig:pot1}
\end{figure}

\begin{figure}
\centering
\includegraphics[width=\linewidth]{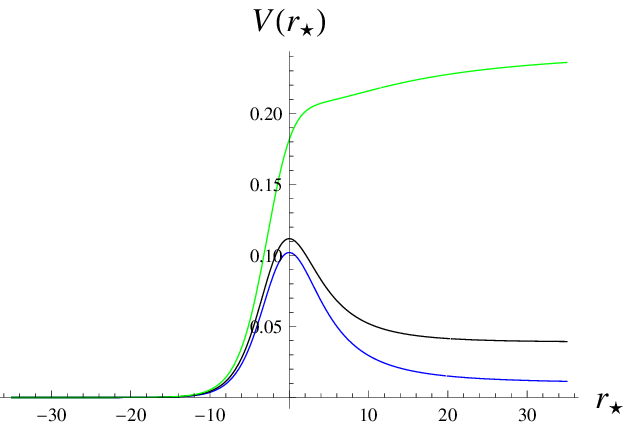}
\caption{Effective potential $V(r_*)$ for the same background as in Fig.~\ref{fig:pot1}, but for $\ell=1$ and $\mu=0.1$ (blue), $\mu=0.2$ (black), and $\mu=0.5$ (green). The angular-momentum term raises the barrier and keeps the single-peak structure more pronounced than in the $\ell=0$ case.}\label{fig:pot2}
\end{figure}

We now consider a test massive scalar field on this fixed background. Its dynamics is governed by the Klein--Gordon equation
\begin{equation}\label{KGg}
\left(\Box-\mu^2\right)\Phi=0,
\end{equation}
where $\mu$ is the mass of the perturbing field. Using the standard decomposition \cite{Carter1968HJ,Carter1968Kerr,Konoplya:2018arm}
\begin{equation}
\Phi(t,r,\theta,\phi)=e^{-i\omega t}Y_{\ell m}(\theta,\phi)\frac{\Psi(r)}{R(r)},
\end{equation}
and introducing the tortoise coordinate
\begin{equation}\label{tortoise}
dr_*\equiv\frac{dr}{f(r)},
\end{equation}
we reduce the radial equation to the Schr\"odinger-like form \cite{Kokkotas:1999bd,Berti:2009kk,Konoplya:2011qq}
\begin{equation}\label{wave-equation}
\frac{d^2 \Psi}{dr_*^2}+(\omega^2-V(r))\Psi=0,
\end{equation}
with the effective potential for $\ell=0,1,2,\ldots$
\begin{equation}\label{potentialScalar}
\begin{split}
V(r)&=f(r)\left(\mu^2+\frac{\ell(\ell+1)}{R^2(r)}\right)\\
&\quad+\frac{1}{R(r)}\frac{d^2 R(r)}{dr_*^2}.
\end{split}
\end{equation}
For the geometry (\ref{metric})--(\ref{metric-functions}) the potential vanishes at the event horizon and approaches $\mu^2$ at spatial infinity, so the massive scalar feels a barrier with a nonzero asymptotic plateau. Increasing $\mu$ raises this plateau and reduces the contrast between the peak and the asymptotic region, while increasing $\ell$ makes the barrier higher through the centrifugal term. The profiles shown in Figs.~\ref{fig:pot1} and \ref{fig:pot2} illustrate these trends for the representative configurations used in our analysis.

\section{Methods}\label{sec:methods}
To compute the quasinormal spectrum we use two complementary techniques. The higher-order WKB approximation improved by Pad\'e resummation provides semi-analytic frequency estimates, while the characteristic time-domain evolution with Prony fitting offers an independent verification of the dominant ringing.

\subsection{WKB approximation}

The WKB approach is especially efficient when the effective potential has a single smooth maximum. In this case the quasinormal frequencies are selected by the boundary conditions
\begin{equation}\label{boundaryconditions}
\Psi(r_*\to\pm\infty)\propto e^{\pm\imo \omega r_*},
\end{equation}
which correspond to purely ingoing waves at the event horizon ($r_*\to-\infty$) and purely outgoing waves at spatial infinity ($r_*\to\infty$).
\par
The WKB construction matches these asymptotic solutions to the Taylor expansion of the potential near its peak. For spherically symmetric backgrounds, the result can be organized as an expansion around the eikonal limit \cite{Konoplya:2019hlu}:
\begin{eqnarray}\label{WKBformula-spherical}
\omega^2&=&V_0+A_2(\K^2)+A_4(\K^2)+A_6(\K^2)+\ldots\\\nonumber&-&\imo \K\sqrt{-2V_2}\left(1+A_3(\K^2)+A_5(\K^2)+A_7(\K^2)\ldots\right),
\end{eqnarray}
while the quasinormal matching condition is
\begin{equation}
\K=n+\frac{1}{2}, \quad n=0,1,2,\ldots,
\end{equation}
where $n$ is the overtone number, $V_0$ is the value of the potential at its maximum, and $V_2$ is the second derivative with respect to the tortoise coordinate at the same point. The correction terms $A_i$ depend on higher derivatives of the potential, and their explicit forms are known up to very high orders: the second and third orders were derived in \cite{Iyer:1986np}, the 4th--6th orders in \cite{Konoplya:2003ii}, and the 7th--13th orders in \cite{Matyjasek:2017psv}. In practice we use high WKB orders together with Pad\'e approximants, which substantially improve convergence for low multipoles and moderate overtones. This approach has been extensively applied to quasinormal modes and grey-body factors in a broad variety of black-hole and compact-object geometries (see, for example, \cite{Abdalla:2005hu,Konoplya:2005sy,Konoplya:2001ji,Konoplya:2006ar,Kokkotas:2010zd,Ishihara:2008re,Konoplya:2018ala,Konoplya:2010vz,Breton:2017hwe,Eniceicu:2019npi,Wongjun:2019ydo,Fernando:2016ftj,Bolokhov:2025egl,Malik:2025erb,Pathrikar:2025gzu,Guo:2020caw,Konoplya:2009hv,Bolokhov:2025aqy,Bolokhov:2025lnt,Skvortsova:2024msa,Malik:2026lfj}).

Because the present work deals with a massive field, it is useful to stress that the same semi-analytic framework has also been widely employed for massive perturbations of black holes and other compact objects, where long-lived modes and oscillatory tails require particularly careful checks against independent time-domain calculations.

\begin{table}
\begin{tabular}{c c c c c}
\hline
a & $\mu$ & WKB16 ($\tilde{m}=8$) & WKB14 ($\tilde{m}=7$) & difference  \\
\hline
$0.2$ & $0$ & $0.110223-0.104792 i$ & $0.110146-0.104450 i$ & $0.230\%$\\
$0.2$ & $0.05$ & $0.110795-0.102774 i$ & $0.110677-0.102335 i$ & $0.301\%$\\
$0.2$ & $0.1$ & $0.112673-0.096808 i$ & $0.115567-0.095714 i$ & $2.08\%$\\
$0.2$ & $0.15$ & $0.114987-0.087658 i$ & $0.115241-0.087064 i$ & $0.447\%$\\
$0.2$ & $0.2$ & $0.123965-0.067932 i$ & $0.114216-0.074534 i$ & $8.33\%$\\
$0.4$ & $0$ & $0.109478-0.104313 i$ & $0.109409-0.103967 i$ & $0.233\%$\\
$0.4$ & $0.05$ & $0.110074-0.102277 i$ & $0.109967-0.101831 i$ & $0.305\%$\\
$0.4$ & $0.1$ & $0.112059-0.096283 i$ & $0.114791-0.095292 i$ & $1.97\%$\\
$0.4$ & $0.15$ & $0.114485-0.087027 i$ & $0.114735-0.086436 i$ & $0.446\%$\\
$0.4$ & $0.2$ & $0.121222-0.068381 i$ & $0.113573-0.073962 i$ & $6.80\%$\\
$0.6$ & $0$ & $0.108264-0.103538 i$ & $0.108206-0.103167 i$ & $0.251\%$\\
$0.6$ & $0.05$ & $0.108900-0.101474 i$ & $0.108812-0.100994 i$ & $0.328\%$\\
$0.6$ & $0.1$ & $0.111110-0.095473 i$ & $0.113683-0.094762 i$ & $1.82\%$\\
$0.6$ & $0.15$ & $0.113674-0.086015 i$ & $0.113882-0.085480 i$ & $0.402\%$\\
$0.6$ & $0.2$ & $0.118468-0.067120 i$ & $0.112595-0.073005 i$ & $6.11\%$\\
$0.8$ & $0$ & $0.106658-0.102505 i$ & $0.106547-0.102221 i$ & $0.206\%$\\
$0.8$ & $0.05$ & $0.107380-0.100402 i$ & $0.107198-0.100023 i$ & $0.286\%$\\
$0.8$ & $0.1$ & $0.109706-0.094017 i$ & $0.112288-0.094665 i$ & $1.84\%$\\
$0.8$ & $0.2$ & $0.114092-0.069721 i$ & $0.111955-0.071132 i$ & $1.92\%$\\
$1.$ & $0$ & $0.104535-0.101138 i$ & $0.104552-0.100903 i$ & $0.163\%$\\
$1.$ & $0.05$ & $0.105263-0.098960 i$ & $0.105290-0.098668 i$ & $0.203\%$\\
$1.$ & $0.1$ & $0.109687-0.092588 i$ & $0.110148-0.091174 i$ & $1.04\%$\\
$1.$ & $0.15$ & $0.111265-0.082667 i$ & $0.112144-0.082382 i$ & $0.666\%$\\
$1.$ & $0.2$ & $0.113653-0.067971 i$ & $0.113378-0.067895 i$ & $0.216\%$\\
\hline
\end{tabular}
\caption{Fundamental quasinormal frequencies ($n=0$) of the massive scalar field with $\ell=0$ in the asymptotically flat regular black-hole background for $M=1$. For each pair of parameters $(a,\mu)$, we compare the 16th-order WKB result with Pad\'e approximant $\tilde{m}=8$ and the 14th-order WKB result with Pad\'e approximant $\tilde{m}=7$; the last column shows the relative difference between the two estimates in percent.}
\end{table}

\begin{table}
\begin{tabular}{c c c c c}
\hline
a & $\mu$ & WKB16 ($\tilde{m}=8$) & WKB14 ($\tilde{m}=7$) & difference  \\
\hline
$0.2$ & $0.05$ & $0.293636-0.096858 i$ & $0.293636-0.096858 i$ & $0\%$\\
$0.2$ & $0.1$ & $0.297011-0.094824 i$ & $0.297011-0.094824 i$ & $0\%$\\
$0.2$ & $0.15$ & $0.302658-0.091382 i$ & $0.302658-0.091382 i$ & $0\%$\\
$0.2$ & $0.2$ & $0.310606-0.086448 i$ & $0.310606-0.086448 i$ & $0\%$\\
$0.2$ & $0.25$ & $0.320887-0.079885 i$ & $0.320887-0.079885 i$ & $0.00003\%$\\
$0.2$ & $0.3$ & $0.333513-0.071493 i$ & $0.333513-0.071493 i$ & $0.\times 10^{\text{-4}}\%$\\
$0.2$ & $0.35$ & $0.348430-0.060998 i$ & $0.348431-0.060998 i$ & $0.0002\%$\\
$0.2$ & $0.4$ & $0.365432-0.048098 i$ & $0.365431-0.048098 i$ & $0.00011\%$\\
$0.2$ & $0.45$ & $0.384939-0.035835 i$ & $0.386689-0.036611 i$ & $0.495\%$\\
$0.4$ & $0$ & $0.291260-0.097149 i$ & $0.291260-0.097149 i$ & $0\%$\\
$0.4$ & $0.05$ & $0.292396-0.096473 i$ & $0.292396-0.096473 i$ & $0\%$\\
$0.4$ & $0.1$ & $0.295811-0.094429 i$ & $0.295811-0.094429 i$ & $0\%$\\
$0.4$ & $0.15$ & $0.301524-0.090972 i$ & $0.301524-0.090972 i$ & $0\%$\\
$0.4$ & $0.2$ & $0.309565-0.086017 i$ & $0.309565-0.086017 i$ & $0\%$\\
$0.4$ & $0.25$ & $0.319964-0.079428 i$ & $0.319964-0.079428 i$ & $0\%$\\
$0.4$ & $0.3$ & $0.332732-0.071005 i$ & $0.332733-0.071005 i$ & $0.0001\%$\\
$0.4$ & $0.35$ & $0.347812-0.060476 i$ & $0.347813-0.060476 i$ & $0.00041\%$\\
$0.4$ & $0.4$ & $0.364984-0.047554 i$ & $0.364979-0.047558 i$ & $0.00188\%$\\
$0.4$ & $0.45$ & $0.385583-0.035752 i$ & $0.387702-0.036248 i$ & $0.562\%$\\
$0.6$ & $0$ & $0.289214-0.096526 i$ & $0.289214-0.096526 i$ & $0\%$\\
$0.6$ & $0.05$ & $0.290372-0.095844 i$ & $0.290372-0.095844 i$ & $0\%$\\
$0.6$ & $0.1$ & $0.293854-0.093785 i$ & $0.293854-0.093785 i$ & $0\%$\\
$0.6$ & $0.15$ & $0.299678-0.090303 i$ & $0.299678-0.090303 i$ & $0\%$\\
$0.6$ & $0.2$ & $0.307872-0.085313 i$ & $0.307872-0.085313 i$ & $0\%$\\
$0.6$ & $0.25$ & $0.318466-0.078681 i$ & $0.318466-0.078681 i$ & $0.\times 10^{\text{-4}}\%$\\
$0.6$ & $0.3$ & $0.331468-0.070209 i$ & $0.331468-0.070208 i$ & $0.\times 10^{\text{-4}}\%$\\
$0.6$ & $0.35$ & $0.346815-0.059624 i$ & $0.346818-0.059625 i$ & $0.001\%$\\
$0.6$ & $0.4$ & $0.364277-0.046660 i$ & $0.364265-0.046673 i$ & $0.00501\%$\\
$0.6$ & $0.45$ & $0.387329-0.035349 i$ & $0.389946-0.035134 i$ & $0.675\%$\\
$0.8$ & $0$ & $0.286438-0.095681 i$ & $0.286438-0.095681 i$ & $0.\times 10^{\text{-4}}\%$\\
$0.8$ & $0.05$ & $0.287627-0.094992 i$ & $0.287627-0.094993 i$ & $0.\times 10^{\text{-4}}\%$\\
$0.8$ & $0.1$ & $0.291201-0.092912 i$ & $0.291201-0.092912 i$ & $0.\times 10^{\text{-4}}\%$\\
$0.8$ & $0.15$ & $0.297178-0.089394 i$ & $0.297177-0.089394 i$ & $0.0001\%$\\
$0.8$ & $0.2$ & $0.305584-0.084357 i$ & $0.305584-0.084357 i$ & $0.0002\%$\\
$0.8$ & $0.25$ & $0.316447-0.077667 i$ & $0.316447-0.077667 i$ & $0.00022\%$\\
$0.8$ & $0.3$ & $0.329771-0.069127 i$ & $0.329771-0.069127 i$ & $0\%$\\
$0.8$ & $0.35$ & $0.345486-0.058469 i$ & $0.345488-0.058478 i$ & $0.00260\%$\\
$0.8$ & $0.4$ & $0.363346-0.045452 i$ & $0.363334-0.045470 i$ & $0.00586\%$\\
$0.8$ & $0.45$ & $0.391212-0.033429 i$ & $0.393991-0.031917 i$ & $0.806\%$\\
$1.$ & $0$ & $0.283009-0.094640 i$ & $0.283009-0.094640 i$ & $0\%$\\
$1.$ & $0.05$ & $0.284238-0.093942 i$ & $0.284238-0.093942 i$ & $0\%$\\
$1.$ & $0.1$ & $0.287929-0.091834 i$ & $0.287929-0.091835 i$ & $0.\times 10^{\text{-4}}\%$\\
$1.$ & $0.15$ & $0.294100-0.088273 i$ & $0.294099-0.088273 i$ & $0.0003\%$\\
$1.$ & $0.2$ & $0.302776-0.083177 i$ & $0.302775-0.083176 i$ & $0.0002\%$\\
$1.$ & $0.25$ & $0.313978-0.076415 i$ & $0.313978-0.076415 i$ & $0.\times 10^{\text{-4}}\%$\\
$1.$ & $0.3$ & $0.327706-0.067792 i$ & $0.327706-0.067791 i$ & $0.00019\%$\\
$1.$ & $0.35$ & $0.343887-0.057042 i$ & $0.343881-0.057056 i$ & $0.00431\%$\\
$1.$ & $0.4$ & $0.362240-0.043968 i$ & $0.362234-0.043978 i$ & $0.00319\%$\\
$1.$ & $0.45$ & $0.397609-0.026875 i$ & $0.399679-0.024201 i$ & $0.849\%$\\
\hline
\end{tabular}
\caption{Fundamental quasinormal frequencies ($n=0$) of the massive scalar field with $\ell=1$ in the asymptotically flat regular black-hole background for $M=1$. The columns labeled WKB16 and WKB14 give the 16th-order and 14th-order WKB results improved by Pad\'e approximants with $\tilde{m}=8$ and $\tilde{m}=7$, respectively, while the last column gives their relative difference in percent.}
\end{table}

\begin{table}
\begin{tabular}{c c c c c}
\hline
a & $\mu$ & WKB16 ($\tilde{m}=8$) & WKB14 ($\tilde{m}=7$) & difference  \\
\hline
$0.2$ & $0.1$ & $0.486154-0.095546 i$ & $0.486154-0.095546 i$ & $0\%$\\
$0.2$ & $0.2$ & $0.495709-0.092256 i$ & $0.495709-0.092256 i$ & $0\%$\\
$0.2$ & $0.3$ & $0.511782-0.086654 i$ & $0.511782-0.086654 i$ & $0\%$\\
$0.2$ & $0.4$ & $0.534609-0.078525 i$ & $0.534609-0.078525 i$ & $0\%$\\
$0.2$ & $0.5$ & $0.564540-0.067463 i$ & $0.564540-0.067463 i$ & $0\%$\\
$0.2$ & $0.6$ & $0.601984-0.052690 i$ & $0.601984-0.052690 i$ & $0\%$\\
$0.2$ & $0.7$ & $0.647029-0.032805 i$ & $0.646987-0.032804 i$ & $0.00651\%$\\
$0.4$ & $0$ & $0.481021-0.096255 i$ & $0.481021-0.096255 i$ & $0\%$\\
$0.4$ & $0.1$ & $0.484224-0.095164 i$ & $0.484224-0.095164 i$ & $0\%$\\
$0.4$ & $0.2$ & $0.493876-0.091861 i$ & $0.493876-0.091861 i$ & $0\%$\\
$0.4$ & $0.3$ & $0.510109-0.086237 i$ & $0.510109-0.086237 i$ & $0\%$\\
$0.4$ & $0.4$ & $0.533157-0.078079 i$ & $0.533157-0.078079 i$ & $0\%$\\
$0.4$ & $0.5$ & $0.563366-0.066981 i$ & $0.563366-0.066981 i$ & $0\%$\\
$0.4$ & $0.6$ & $0.601139-0.052165 i$ & $0.601139-0.052165 i$ & $0\%$\\
$0.4$ & $0.7$ & $0.646541-0.032224 i$ & $0.646488-0.032232 i$ & $0.00829\%$\\
$0.6$ & $0$ & $0.477816-0.095639 i$ & $0.477816-0.095639 i$ & $0\%$\\
$0.6$ & $0.1$ & $0.481073-0.094541 i$ & $0.481073-0.094541 i$ & $0\%$\\
$0.6$ & $0.2$ & $0.490885-0.091215 i$ & $0.490885-0.091215 i$ & $0\%$\\
$0.6$ & $0.3$ & $0.507381-0.085556 i$ & $0.507381-0.085556 i$ & $0\%$\\
$0.6$ & $0.4$ & $0.530794-0.077351 i$ & $0.530794-0.077351 i$ & $0\%$\\
$0.6$ & $0.5$ & $0.561462-0.066194 i$ & $0.561462-0.066194 i$ & $0\%$\\
$0.6$ & $0.6$ & $0.599778-0.051306 i$ & $0.599778-0.051306 i$ & $0.00002\%$\\
$0.6$ & $0.7$ & $0.645752-0.031279 i$ & $0.645681-0.031312 i$ & $0.0121\%$\\
$0.8$ & $0$ & $0.473462-0.094802 i$ & $0.473462-0.094802 i$ & $0\%$\\
$0.8$ & $0.1$ & $0.476793-0.093694 i$ & $0.476793-0.093694 i$ & $0\%$\\
$0.8$ & $0.3$ & $0.503688-0.084630 i$ & $0.503688-0.084630 i$ & $0\%$\\
$0.8$ & $0.4$ & $0.527603-0.076360 i$ & $0.527603-0.076360 i$ & $0\%$\\
$0.8$ & $0.5$ & $0.558902-0.065124 i$ & $0.558902-0.065124 i$ & $0\%$\\
$0.8$ & $0.6$ & $0.597962-0.050140 i$ & $0.597962-0.050140 i$ & $0.00002\%$\\
$0.8$ & $0.7$ & $0.644693-0.030021 i$ & $0.644609-0.030117 i$ & $0.0198\%$\\
$1.$ & $0$ & $0.468076-0.093767 i$ & $0.468076-0.093767 i$ & $0\%$\\
$1.$ & $0.1$ & $0.471501-0.092646 i$ & $0.471501-0.092646 i$ & $0\%$\\
$1.$ & $0.2$ & $0.481815-0.089251 i$ & $0.481815-0.089251 i$ & $0\%$\\
$1.$ & $0.3$ & $0.499139-0.083482 i$ & $0.499139-0.083482 i$ & $0\%$\\
$1.$ & $0.4$ & $0.523689-0.075132 i$ & $0.523689-0.075132 i$ & $0\%$\\
$1.$ & $0.5$ & $0.555782-0.063800 i$ & $0.555782-0.063800 i$ & $0\%$\\
$1.$ & $0.6$ & $0.595772-0.048700 i$ & $0.595772-0.048700 i$ & $0\%$\\
$1.$ & $0.7$ & $0.643409-0.028568 i$ & $0.643368-0.028800 i$ & $0.0367\%$\\
\hline
\end{tabular}
\caption{Fundamental quasinormal frequencies ($n=0$) of the massive scalar field with $\ell=2$ in the asymptotically flat regular black-hole background for $M=1$. The table compares the 16th-order and 14th-order WKB--Pad\'e estimates, corresponding to $\tilde{m}=8$ and $\tilde{m}=7$, and the final column lists the relative difference in percent.}
\end{table}

\begin{table}
\begin{tabular}{c c c c c}
\hline
a & $\mu$ & WKB16 ($\tilde{m}=8$) & WKB14 ($\tilde{m}=7$) & difference  \\
\hline
$0.2$ & $0.1$ & $0.464840-0.292872 i$ & $0.464840-0.292872 i$ & $0\%$\\
$0.2$ & $0.2$ & $0.469683-0.285776 i$ & $0.469683-0.285776 i$ & $0\%$\\
$0.2$ & $0.3$ & $0.477597-0.273731 i$ & $0.477597-0.273730 i$ & $0.00015\%$\\
$0.2$ & $0.4$ & $0.488270-0.256428 i$ & $0.488270-0.256427 i$ & $0.00008\%$\\
$0.2$ & $0.5$ & $0.501096-0.233521 i$ & $0.501096-0.233520 i$ & $0\%$\\
$0.2$ & $0.6$ & $0.515085-0.204885 i$ & $0.515119-0.204757 i$ & $0.0238\%$\\
$0.2$ & $0.7$ & $0.525943-0.172672 i$ & $0.526502-0.173485 i$ & $0.178\%$\\
$0.4$ & $0$ & $0.461316-0.294065 i$ & $0.461316-0.294065 i$ & $0\%$\\
$0.4$ & $0.1$ & $0.462971-0.291711 i$ & $0.462971-0.291711 i$ & $0\%$\\
$0.4$ & $0.2$ & $0.467892-0.284584 i$ & $0.467892-0.284584 i$ & $0\%$\\
$0.4$ & $0.3$ & $0.475937-0.272490 i$ & $0.475936-0.272490 i$ & $0.00012\%$\\
$0.4$ & $0.4$ & $0.486789-0.255120 i$ & $0.486789-0.255119 i$ & $0.00012\%$\\
$0.4$ & $0.5$ & $0.499839-0.232133 i$ & $0.499839-0.232133 i$ & $0\%$\\
$0.4$ & $0.6$ & $0.514095-0.203415 i$ & $0.514124-0.203263 i$ & $0.0281\%$\\
$0.4$ & $0.7$ & $0.525395-0.171375 i$ & $0.526028-0.172042 i$ & $0.166\%$\\
$0.6$ & $0$ & $0.458218-0.292185 i$ & $0.458218-0.292185 i$ & $0\%$\\
$0.6$ & $0.1$ & $0.459916-0.289814 i$ & $0.459916-0.289814 i$ & $0\%$\\
$0.6$ & $0.2$ & $0.464969-0.282637 i$ & $0.464969-0.282637 i$ & $0.00004\%$\\
$0.6$ & $0.3$ & $0.473228-0.270462 i$ & $0.473228-0.270461 i$ & $0.00012\%$\\
$0.6$ & $0.4$ & $0.484377-0.252982 i$ & $0.484377-0.252981 i$ & $0.00017\%$\\
$0.6$ & $0.5$ & $0.497799-0.229865 i$ & $0.497795-0.229861 i$ & $0.00108\%$\\
$0.6$ & $0.6$ & $0.512490-0.201012 i$ & $0.512514-0.200812 i$ & $0.0365\%$\\
$0.6$ & $0.7$ & $0.524573-0.169124 i$ & $0.525367-0.169680 i$ & $0.176\%$\\
$0.8$ & $0$ & $0.454006-0.289629 i$ & $0.454006-0.289629 i$ & $0\%$\\
$0.8$ & $0.1$ & $0.455764-0.287236 i$ & $0.455764-0.287236 i$ & $0\%$\\
$0.8$ & $0.3$ & $0.469557-0.267703 i$ & $0.469556-0.267703 i$ & $0.00014\%$\\
$0.8$ & $0.4$ & $0.481117-0.250074 i$ & $0.481117-0.250073 i$ & $0.00025\%$\\
$0.8$ & $0.5$ & $0.495053-0.226780 i$ & $0.495039-0.226777 i$ & $0.00268\%$\\
$0.8$ & $0.6$ & $0.510338-0.197741 i$ & $0.510357-0.197459 i$ & $0.0516\%$\\
$0.8$ & $0.7$ & $0.523614-0.165967 i$ & $0.524878-0.166488 i$ & $0.249\%$\\
$1.$ & $0$ & $0.448790-0.286467 i$ & $0.448790-0.286467 i$ & $0\%$\\
$1.$ & $0.1$ & $0.450626-0.284045 i$ & $0.450626-0.284045 i$ & $0\%$\\
$1.$ & $0.2$ & $0.456090-0.276713 i$ & $0.456090-0.276713 i$ & $0.00010\%$\\
$1.$ & $0.3$ & $0.465029-0.264287 i$ & $0.465028-0.264286 i$ & $0.00023\%$\\
$1.$ & $0.4$ & $0.477111-0.246470 i$ & $0.477111-0.246468 i$ & $0.00031\%$\\
$1.$ & $0.5$ & $0.491694-0.222958 i$ & $0.491675-0.222959 i$ & $0.00357\%$\\
$1.$ & $0.6$ & $0.507730-0.193690 i$ & $0.507725-0.193263 i$ & $0.0785\%$\\
$1.$ & $0.7$ & $0.523070-0.162093 i$ & $0.525714-0.162445 i$ & $0.487\%$\\
\hline
\end{tabular}
\caption{First-overtone quasinormal frequencies ($n=1$) of the massive scalar field with $\ell=2$ in the asymptotically flat regular black-hole background for $M=1$. As in the previous tables, we compare the 16th-order and 14th-order WKB results improved by Pad\'e approximants with $\tilde{m}=8$ and $\tilde{m}=7$, and quote their relative difference in percent in the last column.}
\end{table}

\begin{table}
\begin{tabular}{c c c c c}
\hline
a & $\mu$ & WKB16 ($\tilde{m}=8$) & WKB14 ($\tilde{m}=7$) & difference  \\
\hline
$0.2$ & $0.1$ & $0.430071-0.505795 i$ & $0.430070-0.505794 i$ & $0.00019\%$\\
$0.2$ & $0.2$ & $0.430384-0.499531 i$ & $0.430383-0.499527 i$ & $0.00056\%$\\
$0.2$ & $0.3$ & $0.430670-0.489206 i$ & $0.430678-0.489200 i$ & $0.00153\%$\\
$0.2$ & $0.4$ & $0.430656-0.475039 i$ & $0.430659-0.475041 i$ & $0.00052\%$\\
$0.2$ & $0.5$ & $0.430025-0.457493 i$ & $0.430025-0.457494 i$ & $0.00009\%$\\
$0.2$ & $0.6$ & $0.428072-0.437297 i$ & $0.428013-0.437281 i$ & $0.00998\%$\\
$0.2$ & $0.7$ & $0.439053-0.439215 i$ & $0.449329-0.439690 i$ & $1.66\%$\\
$0.4$ & $0$ & $0.428156-0.505917 i$ & $0.428156-0.505916 i$ & $0.00020\%$\\
$0.4$ & $0.1$ & $0.428304-0.503809 i$ & $0.428304-0.503808 i$ & $0.00021\%$\\
$0.4$ & $0.2$ & $0.428680-0.497515 i$ & $0.428675-0.497509 i$ & $0.00115\%$\\
$0.4$ & $0.3$ & $0.429072-0.487141 i$ & $0.429073-0.487135 i$ & $0.00093\%$\\
$0.4$ & $0.4$ & $0.429207-0.472911 i$ & $0.429204-0.472919 i$ & $0.00131\%$\\
$0.4$ & $0.5$ & $0.428762-0.455304 i$ & $0.428762-0.455305 i$ & $0.00023\%$\\
$0.4$ & $0.6$ & $0.426985-0.435101 i$ & $0.426885-0.435083 i$ & $0.0166\%$\\
$0.4$ & $0.7$ & $0.442642-0.438716 i$ & $0.449276-0.436813 i$ & $1.11\%$\\
$0.6$ & $0$ & $0.425232-0.502688 i$ & $0.425232-0.502686 i$ & $0.00030\%$\\
$0.6$ & $0.1$ & $0.425414-0.500564 i$ & $0.425415-0.500562 i$ & $0.00031\%$\\
$0.6$ & $0.2$ & $0.425894-0.494222 i$ & $0.425886-0.494216 i$ & $0.00154\%$\\
$0.6$ & $0.3$ & $0.426458-0.483768 i$ & $0.426457-0.483766 i$ & $0.00044\%$\\
$0.6$ & $0.4$ & $0.426836-0.469439 i$ & $0.426830-0.469450 i$ & $0.00213\%$\\
$0.6$ & $0.5$ & $0.426709-0.451730 i$ & $0.426705-0.451740 i$ & $0.00172\%$\\
$0.6$ & $0.6$ & $0.425235-0.431503 i$ & $0.425074-0.431491 i$ & $0.0267\%$\\
$0.6$ & $0.7$ & $0.445268-0.434427 i$ & $0.449013-0.432031 i$ & $0.715\%$\\
$0.8$ & $0$ & $0.421248-0.498295 i$ & $0.421249-0.498292 i$ & $0.00056\%$\\
$0.8$ & $0.1$ & $0.421479-0.496148 i$ & $0.421480-0.496145 i$ & $0.00057\%$\\
$0.8$ & $0.3$ & $0.422906-0.479178 i$ & $0.422905-0.479176 i$ & $0.00044\%$\\
$0.8$ & $0.4$ & $0.423622-0.464709 i$ & $0.423610-0.464728 i$ & $0.00361\%$\\
$0.8$ & $0.5$ & $0.423940-0.446867 i$ & $0.423928-0.446901 i$ & $0.00586\%$\\
$0.8$ & $0.6$ & $0.422884-0.426596 i$ & $0.422664-0.426598 i$ & $0.0367\%$\\
$0.8$ & $0.7$ & $0.445823-0.427903 i$ & $0.448802-0.424971 i$ & $0.676\%$\\
$1.$ & $0$ & $0.416304-0.492855 i$ & $0.416304-0.492848 i$ & $0.00112\%$\\
$1.$ & $0.1$ & $0.416596-0.490680 i$ & $0.416596-0.490672 i$ & $0.00117\%$\\
$1.$ & $0.2$ & $0.417405-0.484192 i$ & $0.417394-0.484177 i$ & $0.00291\%$\\
$1.$ & $0.3$ & $0.418513-0.473490 i$ & $0.418510-0.473486 i$ & $0.00078\%$\\
$1.$ & $0.4$ & $0.419656-0.458844 i$ & $0.419635-0.458864 i$ & $0.00461\%$\\
$1.$ & $0.5$ & $0.420548-0.440843 i$ & $0.420528-0.440909 i$ & $0.0113\%$\\
$1.$ & $0.6$ & $0.419996-0.420508 i$ & $0.419754-0.420523 i$ & $0.0409\%$\\
$1.$ & $0.7$ & $0.445780-0.419267 i$ & $0.449860-0.413482 i$ & $1.16\%$\\
\hline
\end{tabular}
\caption{Second-overtone quasinormal frequencies ($n=2$) of the massive scalar field with $\ell=2$ in the asymptotically flat regular black-hole background for $M=1$. As in the previous tables, we compare the 16th-order and 14th-order WKB results improved by Pad\'e approximants with $\tilde{m}=8$ and $\tilde{m}=7$, and quote their relative difference in percent in the last column.}
\end{table}

\subsection{Time-domain integration}

To complement the WKB analysis, we perform a characteristic integration of the wave equation in the time domain. Introducing null coordinates $u=t-r_*$ and $v=t+r_*$, we evolve the perturbation by means of the Gundlach--Price--Pullin discretization scheme \cite{Gundlach:1993tp}
\begin{eqnarray}
\Psi\left(N\right)&=&\Psi\left(W\right)+\Psi\left(E\right)-\Psi\left(S\right)\nonumber\\
&&- \Delta^2V\left(S\right)\frac{\Psi\left(W\right)+\Psi\left(E\right)}{8}+{\cal O}\left(\Delta^4\right),\label{Discretization}
\end{eqnarray}
where $N\equiv\left(u+\Delta,v+\Delta\right)$, $W\equiv\left(u+\Delta,v\right)$, $E\equiv\left(u,v+\Delta\right)$, and $S\equiv\left(u,v\right)$. Starting from prescribed initial data on the null surfaces, this scheme reconstructs the full waveform, including the prompt response, the quasinormal ringing stage, and the subsequent tail. It has been used in a large number of black-hole and wormhole studies and provides a robust independent check of frequency-domain calculations \cite{Dubinsky:2025fwv,Bolokhov:2024ixe,Konoplya:2013sba,Malik:2024nhy,Konoplya:2024hfg,Skvortsova:2024atk,Malik:2023bxc,Varghese:2011ku,Skvortsova:2023zmj,Skvortsova:2024wly,Bolokhov:2023dxq,Dubinsky:2024nzo,Abdalla:2012si,Bolokhov:2026eqf,Arbelaez:2026eaz,Arbelaez:2025gwj}.

\begin{figure}
\centering
\includegraphics[width=\linewidth]{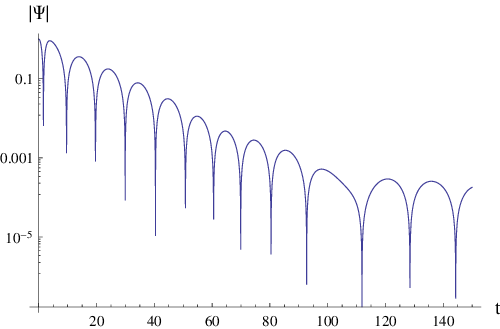}
\caption{Time-domain profile $|\Psi|$ for the massive scalar field in the asymptotically flat regular black-hole background with $\ell=1$, $a=0.5$, $\mu=0.2$, and $M=1$. The waveform is shown on a semi-logarithmic scale as a function of time $t$; the damped oscillatory ringdown is followed by the onset of the late-time regime. The Prony method gives $\omega = 0.3084 - 0.0855 i$, while the WKB method gives $\omega = 0.308797 - 0.085697 i$. Thus, the real and imaginary parts agree to within $0.13\%$ and $0.23\%$, respectively.}
\label{fig:tdprofile}
\end{figure}

A representative time-domain profile for the massive scalar perturbation with $\ell=1$, $a=0.5$, $\mu=0.2$, and $M=1$ is shown in Fig.~\ref{fig:tdprofile}. The damped oscillatory ringdown is clearly visible on the logarithmic plot and provides the intermediate-time window from which the dominant quasinormal frequencies can be extracted.

\begin{figure}
\centering
\includegraphics[width=\linewidth]{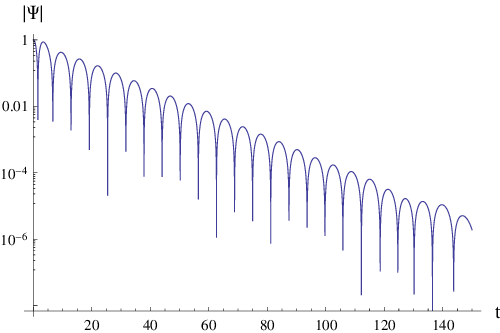}
\caption{Time-domain profile $|\Psi|$ for the massive scalar field in the asymptotically flat regular black-hole background with $\ell=2$, $a=0.6$, $\mu=0.3$, and $M=1$. The waveform is shown on a semi-logarithmic scale as a function of time $t$; the damped oscillatory ringdown is followed by the onset of the late-time regime. The Prony method gives $\omega = 0.507416 - 0.0855559 i$, while the WKB method gives $\omega = 0.507381 - 0.085556 i$. Thus, the real and imaginary parts agree to within $0.0069\%$ and $0.00012\%$, respectively.}
\label{fig:tdprofile2}
\end{figure}

Another representative time-domain profile for the configuration $\ell=2$, $a=0.6$, $\mu=0.3$, and $M=1$ is shown in Fig.~\ref{fig:tdprofile2}. In this case the Prony extraction reproduces the corresponding WKB value from Table~III with excellent accuracy, further illustrating the consistency between the time-domain and frequency-domain approaches.

\begin{figure}
\centering
\includegraphics[width=\linewidth]{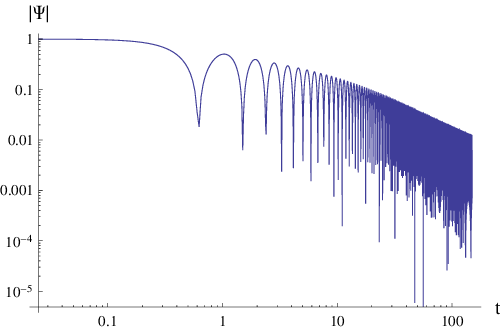}
\caption{Logarithmic time-domain profile $|\Psi|$ for the massive scalar field in the asymptotically flat regular black-hole background with $\ell=0$, $a=0.6$, $\mu=5$, and $M=1$. The intermediate and asymptotic tails begin dominating the signal already at relatively early times, so that the long lived modes are suppressed by oscillatory tails with power-law envelope.}
\label{fig:tdprofile3}
\end{figure}

To extract the quasinormal frequencies from the numerical signal we apply the Prony method to the intermediate stage of the profile, where the prompt transient has already decayed while the late-time tail is still subdominant. In this approach the waveform is approximated by a finite sum of damped exponentials, $\Psi(t)\approx \sum_{k=1}^{p} C_k e^{-i\omega_k t}$, and the complex frequencies are recovered from a linear prediction problem for a discrete set of time samples. By varying the fitting window and the number of exponentials, one can check the stability of the dominant mode and compare the extracted frequency with the WKB result. This fitting procedure is standard in black-hole spectroscopy and is widely used as a practical way of reading off the ringing frequencies from time-domain data \cite{Berti:2009kk,Konoplya:2011qq}.

In the present work, the WKB--Pad\'e method and the time-domain evolution with Prony extraction play complementary roles: the former provides efficient semi-analytic estimates of the frequencies, while the latter independently validates the dominant quasinormal ringing seen in the numerical profiles.

\section{Quasinormal modes}

The tables show that, except for the lowest multipole at the largest field masses, the internal WKB--Pad\'e uncertainty is much smaller than the physical shift produced by the regularity parameter $a$. For example, for the fundamental $\ell=1$ mode at $\mu=0.3$, changing $a$ from $0.2$ to $1$ shifts $\mathrm{Re}(\omega)$ by about $1.7\%$ and $|\mathrm{Im}(\omega)|$ by about $5.2\%$, while the relative difference between the 16th-order and 14th-order WKB--Pad\'e estimates stays below $2\times 10^{-4}\%$. A similar hierarchy holds for the $\ell=2$ fundamental mode and for the first two overtones: the variation with $a$ is typically of order a few percent, whereas the WKB consistency check is usually in the range $10^{-4}\%$--$10^{-1}\%$. Therefore the dependence on the regularity scale is a genuine physical effect and not a numerical artifact. The only visible exception is the $\ell=0$ sector near the largest masses, where the WKB spread reaches a few percent, so those entries must be treated with additional caution.

The dependence of the spectrum on the parameters is also clear. For fixed $n$ and $\ell$, increasing the field mass $\mu$ increases $\mathrm{Re}(\omega)$ and at the same time decreases $|\mathrm{Im}(\omega)|$, so the modes oscillate faster but decay more slowly as the field becomes heavier. Increasing the regularity parameter $a$ generally lowers both the oscillation frequency and the damping rate, meaning that the regular geometry makes the ringing slightly softer and longer lived. The multipole number $\ell$ mainly raises the oscillation frequency through the centrifugal part of the effective potential, whereas the overtone number $n$ primarily controls the damping: the first and second overtones decay much faster than the fundamental mode, even when their real parts remain of the same order.

\begin{figure}
\centering
\includegraphics[width=\linewidth]{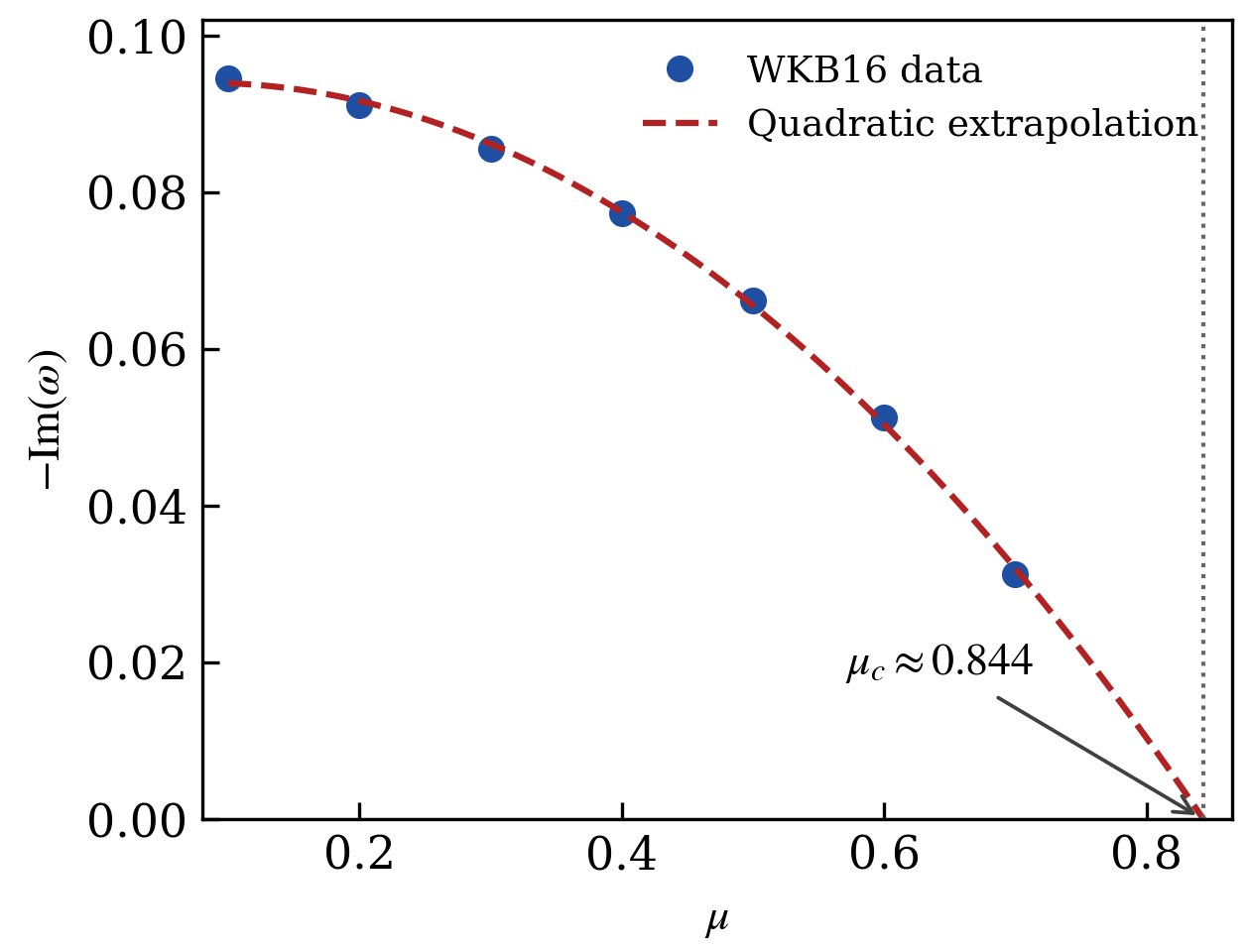}
\caption{Damping rate $-\mathrm{Im}(\omega)$ for the fundamental massive scalar mode with $\ell=1$ and $a=0.6$ in the asymptotically flat regular black-hole background with $M=1$, plotted as a function of the field mass $\mu$. The points show the WKB16 data from Table~II, while the dashed curve is a quadratic extrapolation of these values. The vertical dotted line marks the critical value $\mu_c\approx 0.844$ at which the extrapolated damping rate vanishes, signaling the onset of a quasiresonant regime.}
\label{fig:dampingrate}
\end{figure}

The monotonic decrease of the damping rate with increasing $\mu$ indicates the approach to quasiresonances, that is, arbitrarily long-lived modes with vanishing damping \cite{Ohashi:2004wr,Konoplya:2004wg,Konoplya:2005hr}. For the fundamental mode with $\ell=1$, $a=0.6$, and $M=1$, the WKB16 data shown in Fig.~\ref{fig:dampingrate} are well described by a quadratic extrapolation, which crosses the horizontal axis at $\mu_c\approx 0.844$. This value should be understood as an extrapolated estimate rather than an exact threshold, but it gives a clear indication that the spectrum tends toward a regime of very long-lived oscillations as the field mass grows.

At the same time, one should not expect these extremely long-lived modes to appear as an arbitrarily extended ringing stage in the time-domain profiles. The Prony method can be applied only in the intermediate regime, after the prompt response and before the onset of the late-time tail. For massive fields, however, the asymptotic signal is governed by oscillatory tails with a power-law envelope \cite{Jing:2004zb,Koyama:2001qw,Moderski:2001tk,Rogatko:2007zz}, and quasinormal modes do not form a complete set. Consequently, when $|\mathrm{Im}(\omega)|$ becomes very small, the waveform is overtaken by the asymptotic tail before the corresponding quasinormal mode can dominate as a clean exponentially damped signal. Therefore the absence of a visibly dominant ultra-long-lived ringdown stage in the time-domain evolution does not contradict the frequency-domain indication of quasiresonances.

Although in the present work we focus on quasinormal frequencies, the same data can also be used to estimate the grey-body factors. Following the correspondence established in Eqs.~(3.5)--(4.6) of Ref.~\cite{Konoplya:2024gbf,Konoplya:2024vuj}, one first writes the partial grey-body factor in the standard WKB form $\Gamma_{\ell}(\Omega)=|T_{\ell}(\Omega)|^2=\left(1+e^{2\pi i\K_{\ell}(\Omega)}\right)^{-1}$, where $\Omega$ is the real scattering frequency and $\K_{\ell}(\Omega)$ is the WKB quantity defined by the scattering problem. In the eikonal regime, $\K_{\ell}(\Omega)$ can be reconstructed from the fundamental quasinormal mode alone through Eq.~(3.5) of \cite{Konoplya:2024gbf}, while beyond the eikonal limit the improved approximation uses the fundamental mode $\omega_{\ell 0}$ together with the first overtone $\omega_{\ell 1}$; in particular, Eq.~(4.6) of \cite{Konoplya:2024gbf,Konoplya:2024vuj} expresses the grey-body factors through these two dominant frequencies. Therefore, once $\omega_{\ell 0}$ and $\omega_{\ell 1}$ are known, one obtains a practical semi-analytic estimate of the transmission probability without solving the scattering problem separately.
It should be noted that this correspondence has recently been tested and applied in a number of works \cite{Lutfuoglu:2025mqa,Lutfuoglu:2025kqp,Lutfuoglu:2026gey,Konoplya:2010vz,Lutfuoglu:2025eik,Lutfuoglu:2026uzy,Dubinsky:2024vbn,Malik:2025qnr,Malik:2024wvs}, demonstrating good accuracy already at moderate values of $\ell$. The correspondence remains valid provided that the effective potential has a single-barrier shape; it may break down in the presence of a double-well structure \cite{Konoplya:2025hgp} or when specific higher-curvature corrections modify the centrifugal term in the effective potential, potentially leading to catastrophic instabilities \cite{Konoplya:2017ymp,Konoplya:2017lhs,Takahashi:2010gz,Dotti:2004sh,Gleiser:2005ra}.

In order to access the regime of quasi-resonances, i.e., modes with arbitrarily small damping rates, one must employ a numerical method based on a convergent procedure, such as the Leaver method \cite{Leaver:1985ax}. However, the application of this approach requires casting the metric into a rational parametrized form, as, for example, in \cite{Konoplya:2020hyk}.

\section{Conclusions}

In this paper we have studied the quasinormal spectrum of a massive test scalar field in the asymptotically flat regular black-hole geometry supported by a phantom Dirac--Born--Infeld scalar. Using high-order WKB approximation improved by Pad\'e resummation, together with characteristic time-domain evolution and Prony extraction, we calculated the fundamental mode and the first two overtones for representative values of the regularity parameter $a$ and the field mass $\mu$. For $\ell\geq 1$ and moderate masses, the agreement between the 16th-order and 14th-order WKB--Pad\'e results is considerably better than the actual shift of the spectrum caused by the nonzero regularity scale, which shows that the observed dependence on $a$ is a robust physical effect. The time-domain profiles confirm this picture and provide an independent check of the dominant ringing frequencies.

The main physical trends are clear. Increasing the field mass raises the oscillation frequency and decreases the damping rate, whereas increasing the regularity parameter generally lowers both $\mathrm{Re}(\omega)$ and $|\mathrm{Im}(\omega)|$, making the ringing softer and longer lived. At fixed $a$ and $\mu$, higher multipoles oscillate faster, while higher overtones are substantially more strongly damped. We have also shown that the monotonic decrease of $|\mathrm{Im}(\omega)|$ with increasing $\mu$ points to the onset of quasiresonances, i.e. arbitrarily long-lived modes. At the same time, the time-domain analysis clarifies why these very long-lived modes are difficult to observe directly as a clean late-time ringdown: for massive fields, oscillatory tails with power-law envelope dominate the asymptotic signal (see fig. \ref{fig:tdprofile3}) before an ultra-long-lived quasinormal mode can form an isolated exponentially damped stage.

These results indicate that the regularity scale of the DBI-supported geometry leaves a visible imprint on black-hole spectroscopy and that massive perturbations provide a sensitive probe of this effect. A natural continuation of the present work is to extend the analysis to electromagnetic, Dirac, and gravitational perturbations, as well as to the extremal-remnant and horizonless branches of the same solution. Another natural direction is a dedicated study of grey-body factors, scattering, and late-time tails in the quasiresonant regime, where the interplay between long-lived modes and asymptotic tails is expected to be especially rich.

\begin{acknowledgments}
The author would like to acknowledge Roman Konoplya for useful discussions.
\end{acknowledgments}

\bibliographystyle{apsrev4-1}
\bibliography{bibliography}
\end{document}